\documentstyle[12pt]{article}
 \makeatletter
 \def\leqq{\mathrel{\mathpalette\gl@align<}}
 \def\geqq{\mathrel{\mathpalette\gl@align>}}
 \def\gl@align#1#2{\lower.6ex\vbox{\baselineskip\z@skip\lineskip\z@
     \ialign{$\m@th#1\hfil##\hfil$\crcr#2\crcr=\crcr}}}
 \makeatother

 \makeatletter
 \def\sileqq{\mathrel{\mathpalette\gs@align<}}
 \def\sigeqq{\mathrel{\mathpalette\gs@align>}}
 \def\gs@align#1#2{\lower.6ex\vbox{\baselineskip\z@skip\lineskip\z@
     \ialign{$\m@th#1\hfil##\hfil$\crcr#2\crcr\sim\crcr}}}
 \makeatother
\begin{document}
\hbadness=10000
\hbadness=10000
\begin{titlepage}
\nopagebreak
\begin{flushright}
{\normalsize
HIP-1999-45/TH\\
KANAZAWA-99-16\\
July, 1999}\\
\end{flushright}
\vspace{0.5cm}
\begin{center}

{\large \bf  $F$-term Inflation in M-theory with Five-branes}

\vspace{0.8cm}

{ S.M. Harun-or-Rashid$^a$}, 
{ Tatsuo Kobayashi$^{a,b}$
}  
and 
{ Hitoshi Shimabukuro$^c$
}

\vspace{0.5cm}
$^a$ Department of Physics, 
FIN-00014 
University of Helsinki, Finland  \\
$^b$ Helsinki Institute of Physics
FIN-00014 University of Helsinki, Finland \\
and\\
$^c$ Institute for Theoretical Physics, Kanazawa University\\
Kanazawa, 920-1192, Japan\\

\end{center}
\vspace{0.5cm}

\nopagebreak

\begin{abstract}
We study $F$-term inflation in M-theory with and 
without five-brane moduli fields.
We show the slow rolling condition is not satisfied in 
M-theory without five-brane moduli fields, but 
it can be satisfied in the case with non-vanishing 
$F$-terms of five-brane moduli fields.
\end{abstract}
\vfill
\end{titlepage}
\pagestyle{plain}
\newpage
\def\thefootnote{\fnsymbol{footnote}}

\section{Introduction}

Cosmological inflation can solve the flatness and 
horizon problems of the universe \cite{review}.
During inflation, the vacuum energy $V_0 \equiv <V>$ takes a value 
$V_0 =3H^2M^2$, where $M$ is the reduced Planck scale and $H$ is 
expected to be of $O(10^{13}\sim 10^{14})$ GeV.
The successful inflation scenario also requires the so-called 
slow-rolling condition for the inflaton field $\Phi$,  
that is, the parameters $\varepsilon_{in} = {1 \over 2}M^2(V'/V)^2$ and 
$\eta = M^2V''/V$ should be suppressed, 
\begin{equation}
\varepsilon_{in} << 1, \qquad \eta << 1.
\label{flatV}
\end{equation}

Within the framework of supergravity, 
the scalar potential $V$ includes the $F$-term and $D$-term contributions,
\begin{equation}
V=F^I\bar F^{\bar J}\partial_I \partial_{\bar J}K - 3e^GM^4
+{(D{\rm -term})}.
\end{equation}
Here $K$ denotes the K\"ahler potential and $G$ is obtained as 
$G=KM^{-2}+\ln(|W|^2M^{-6})$, where $W$ is the superpotential.
Thus, the nonvanishing vacuum energy $V_0$ can be induced by 
nonvanishing $F$-terms and/or $D$-terms.
That implies supersymmetry breaking.

Here we consider the case that $F$-terms contributions are dominant 
within the framework of string-inspired 
supergravity \cite{st-inf1,st-inf2,st-inf3}, 
although the $D$-term inflation \cite{D-inf} induced 
by the anomalous $U(1)$ symmetry \cite{au1}
is also another interesting possibility.
Within the framework of  superstring theory, 
the dilaton and moduli fields can be candidates for the inflation  
field \cite{st-inf1}.
However, here we consider other matter fields 
as candidates for the inflaton in hybrid inflation, 
which is driven by the vacuum energy due to non-vanishing $F$-terms 
of the dilaton and moduli fields.

In general, non-vanishing $F$-terms, however, induce a seizable soft 
scalar mass $m_\Phi$ for the inflaton field $\Phi$, and that spoils 
the flatness condition (\ref{flatV}) of $V$ for $\Phi$.
This problem has been discussed within the framework of weakly 
coupled superstring theory.
In Ref.\cite{st-inf4} it has been shown that 
we have $m^2_\Phi =0$ for the field with 
the modular weight $n=-3$ and the vacuum energy $V_0$ driven by 
the $F$-term of the moduli field $T$ in weakly coupled heterotic string 
theory without nonperturbative K\"ahler potential.

In this paper we consider the condition $m^2_\Phi =0$ for $V_0 \neq 0$ 
within the framework of strongly coupled heterotic sting theory, 
M-theory \cite{M-theory}.
We take into account effects due to five-brane moduli fields.

\section{M-theory without five-brane}

First we consider the case without five-brane.
The K\"ahler potential $K$ is obtained \cite{LOW}, 
\begin{eqnarray}
K&=&-\log(S+\bar{S})-3\log(T+\bar{T})+\left(\frac{3}{T+\bar{T}}
   +\frac{\alpha}{S+\bar{S}}\right)|\Phi|^2 \label{K},
\end{eqnarray}
where $\alpha={1}/{(8\pi^2)}\int \omega \wedge \left[tr(F
\wedge F)-\frac{1}{2}tr(R \wedge R)\right]$.
The fields $S$ and $T$ denote the dilation field and moduli field.
We assume that the $F$-terms of $S$ and $T$ contribute to $V_0$.
Then we parameterize the $F$-terms, 
\begin{eqnarray}
F^S&=&\sqrt{3}m_{3/2}C(S+\bar{S})\sin\theta  \label{Fs},\\
F^T&=&m_{3/2}C(T+\bar{T})\cos\theta ,\label{Ft}
\end{eqnarray}
where $C^2=1+V_0/(3m_{3/2}^2M^2)$ and $m_{3/2}=e^GM^2$. 
We have neglected the CP phases of $F$-terms.
Using this parametrizatioin, the soft scalar mass $m_\Phi^2$ 
is obtained \cite{CKM}, 
\begin{eqnarray}
m_\Phi^2&=&V_0M^{-2}+m^2_{3/2}-\frac{3C^2m^2_{3/2}}{3(S+\bar{S})
+\alpha(T+\bar{T})} 
     \nonumber\\
&{}& \times 
 \left\{
    \alpha(T+\bar{T})\left(2-\frac{\alpha(T+\bar{T})}{3(S+\bar{S})
       +\alpha(T+\bar{T})}\right)\sin^2\theta \right.  \nonumber\\
&{}& \left. +(S+\bar{S})\left(2-\frac{3(S+\bar{S})}{3(S+\bar{S})
       +\alpha(T+\bar{T})}\right)\cos^2\theta \right.\nonumber\\
&{}& \left. -\frac{2\sqrt{3}\alpha(T+\bar{T})(S+\bar{S})}{3(S+\bar{S})
       +\alpha(T+\bar{T})}\sin\theta\cos\theta
 \right\} \label{ms}.
\end{eqnarray}
We take the limit $C^2>>1$, i.e. $V_0 >> 3m_{3/2}^2M^2$, 
and investigate the condition for 
$m_\Phi^2<<O(H^2=V_0M^{-2})$.
It is obvious that we have 
$m_\Phi^2 =O(H^2)$ in most of the parameter space.
However, the condition $m_\Phi^2 =0$ is realized for the following 
case:
\begin{eqnarray}
\cos \theta =0, \qquad 
{(S+\bar{S}) \over \alpha(T+\bar{T})} =0.
\label{cond0} 
\end{eqnarray}

Unfortunately, the solution (\ref{cond0}) is not a realistic 
solution from the viewpoint of M-theory.
In M-theory, we have two sectors, which are usually called the 
observable sector and the hidden sector.
The sector including the scalar field $\Phi$ has the gauge kinetic 
function $f$ \cite{g-kin}, 
\begin{equation}
f=S+\alpha T,
\label{g-inf}
\end{equation}
and the other sector has the gauge kinetic function 
$f'$, 
\begin{equation}
f'=S+\alpha' T.
\label{g-other}
\end{equation}
The gauge couplings of these sectors $g$ and $g'$ are obtained as 
$g^{-2}=Re(f)$ and $g'^{-2}=Re(f')$.
The coefficients $\alpha$ and $\alpha'$ should satisfy 
\begin{equation}
\alpha +\alpha'=0.
\label{const1}
\end{equation}
It is impossible that the both of $g^{-2}$ and $g'^{-2}$ take 
positive values for the condition (\ref{cond0}), 
i.e. 
$(S+\bar S)<<\alpha (T+\bar T)$.
Thus, the condition $m_\Phi^2 << O(H^2)$ can not be realized 
in the $F$-term inflation of M-theory without five-brane.

\section{M-theory with five-branes}

Next we consider M-theory with five-brane moduli fields 
$Z^n$ \cite{5brane}, 
whose vacuum expectation values provide with 
positions $z^n$ of the corresponding five-branes along the orbifold 
$S^1/Z_2$.
The moduli K\"ahler potential $K_{mod}$ and the 
K\"ahler potential of the scalar field $\Phi$ $K_\Phi$ are obtained 
\begin{eqnarray}
K_{mod}&=&-ln(S+\bar{S})-3ln(T+\bar{T})+K_5 \label{Kmod}, \\
K_{\Phi}&=&\left(\frac{3}{(T+\bar{T})}+\frac{\epsilon\zeta}
            {(S+\bar{S})}\right)|\Phi|^2, 
\end{eqnarray}
where 
\begin{eqnarray}
\zeta=\beta+\sum_{n=1}^{N}(1-z^n)^2\beta^{(n)}\label{zeta}.
\end{eqnarray}
Here $K_5$ denotes the K$\ddot{\mbox{a}}$hler potential for $Z^n$, 
$\epsilon$ is the expansion parameter 
$\epsilon=\left({\kappa}/{4\pi}\right)^{2/3}{2\pi^2\rho}/{V^{2/3}}$, 
$\beta^{(n)}$  is a magnetic charge on the each
5-brane and   $\beta$ is the instanton number on the
sector boundary including the inflaton $\Phi$.
The gauge kinetic functions of the sector including the inflaton 
$\Phi$ and the other sector, $f$ and $f'$, are obtained 
in the same forms as eqs.(\ref{g-inf}) and (\ref{g-other}) except 
changing $\alpha$ and $\alpha'$,
\begin{eqnarray}
\alpha&=&\epsilon \left(\beta
           +\sum_{n=1}^{N}(1-z^n)^2\beta^{(n)}\right),
\label{5gfun1}
\nonumber \\
\alpha'&=&\epsilon \left(\beta'
            +\sum_{n=1}^{N}(z^n)^2\beta^{(n)}\right) \label{5gfun2}.
\end{eqnarray}
We have the constraint 
\begin{eqnarray}
\beta+\beta'+\sum_{n=1}^{N}\beta^{(n)}=0 \label{coho}.
\end{eqnarray}
The condition  (\ref{const1}) is relaxed by $\beta^{(n)}$.
Thus the solution (\ref{cond0}) could be realistic and both $g^{-2}$ 
and $g'^{-2}$ could be positive if 
both $\alpha$ and $\alpha'$ are positive, 
or if one of them is positive and the other is suppressed enough, e.g. 
$\alpha >0$ and $\alpha' \approx 0$.

Furthermore, the $F$-term of the five-brane moduli $Z^n$ can also 
contribute to $V_0$ and $m_\Phi^2$.
That changes the situation and could give a realistic solution  
even for $\alpha \alpha' <0$.
In this section, we consider effects of the $F$-term of $Z^n$.

For simplicity, we assume that there is only one relevant 
5-brane moduli $Z$ and its K\"ahler potential is a function of only 
$(Z+\bar Z)$.
Now we can parameterize each $F$-term as follows
\begin{eqnarray}
F^S&=&\sqrt{3}m_{3/2}C(S+\bar{S})\sin\theta \sin\phi, \nonumber \\
F^T&=&m_{3/2}C(T+\bar{T})\cos\theta \sin\phi, \nonumber \\
F^Z&=&\sqrt{\frac{3}{K_{5,Z\bar Z}}}m_{3/2}C\cos\phi.
\end{eqnarray}
In addition, we fix $z$ and $\beta^{(1)}$, e.g. 
\begin{eqnarray}
z=\frac{1}{2},\hspace{1cm} \beta^{(1)}=\frac{4}{3}\beta \label{asmp2}.
\end{eqnarray}
Note that in this case we have 
\begin{eqnarray}
2Re(f) &=& (S+\bar{S})+\epsilon\zeta(T+\bar{T}),
\label{5const} \\
2Re(f') &=& (S+\bar{S})-\frac{3}{2}\epsilon\zeta(T+\bar{T}),
\label{5const2}
\end{eqnarray}
i.e. $\alpha \alpha' <0$, where $\alpha = \epsilon\zeta$ 
and $\alpha' = -3\epsilon\zeta/2$.

In the limit $C>>1$, we obtain 
$m_\Phi^2$ \cite{5b-soft} \footnote{See also Ref.\cite{5b-soft2}.}, 
\begin{eqnarray}
m^2_\Phi&=&3m_{3/2}^2C^2
\left\{
          1-\Bigl[1-\frac{9(S+\bar{S})^2}{\Bigl(3(S+\bar{S})
          +\epsilon\zeta(T+\bar{T})\Bigr)^2}\Bigr]\sin^2\theta \sin^2\phi 
\right.
\nonumber \\
   &-&{1 \over 3}\Bigl[1-\Bigl(\frac{\epsilon\zeta(T+\bar{T})}
          {(3(S+\bar{S})+\epsilon\zeta(T+\bar{T})}\Bigr)^2\Bigr]
           \cos^2\theta \sin^2\phi \nonumber \\
   &-&\frac{\epsilon\zeta(T+\bar{T})}{K_{5,Z \bar Z}
          \Bigl(3(S+\bar{S})+\epsilon\zeta(T+\bar{T})\Bigr)}
        \Bigl[2-\frac{\epsilon\zeta(T+\bar{T})}
      {(3(S+\bar{S})+\epsilon\zeta(T+\bar{T})}\Bigr]\cos^2\phi \nonumber \\
   &+&\frac{2\sqrt{3}(S+\bar{S})\epsilon\zeta(T+\bar{T})}
       {\Bigl(3(S+\bar{S})+\epsilon\zeta(T+\bar{T})\Bigr)^2}
           \sin\theta \cos\theta \sin^2\phi \nonumber \\
   &-&\frac{2\sqrt{3}(S+\bar{S})\epsilon\zeta(T+\bar{T})}
              {\Bigl(3(S+\bar{S})+\epsilon\zeta(T+\bar{T})\Bigr)^2}
            \sqrt{\frac{3}{K_{5,Z \bar Z}}}\sin\theta \sin\phi \cos\phi 
\nonumber \\
   &+&\frac{2\epsilon\zeta(T+\bar{T})}{3(S+\bar{S})+\epsilon\zeta(T+\bar{T})}
            \Bigl[1-\frac{\epsilon\zeta(T+\bar{T})}
         {(3(S+\bar{S})+\epsilon\zeta(T+\bar{T})}\Bigr]\nonumber \\
&\times&
\left.   
        \sqrt{\frac{1}{3 K_{5,Z \bar Z}}}\cos\theta \sin\phi \cos\phi 
\right\}.
\label{5mass}
\end{eqnarray}

The region with $\cos \theta =0$ in eq.(6) 
gives a solution for $m_\Phi^2=0$ 
without five-brane, 
although it is not realistic because of negative gauge couplings squared.
Hence, let us consider the condition $m_\Phi^2=0$ for 
$\cos \theta =0$ in eq.(\ref{5mass}) at first.
For $K_{5,Z \bar Z}=1$, the equation becomes simple and 
the two solutions for $m_\Phi^2=0$ are obtained 
\begin{equation}
{(S+\bar{S}) \over \alpha(T+\bar{T})} =0,\qquad
{(S+\bar{S}) \over \alpha(T+\bar{T})} ={2 \over 3} \sin \phi \cos \phi.
\end{equation}
While the former corresponds to the solution (\ref{cond0})  
in the case without 
five-brane, the latter is a new solution.
However, the latter is not realistic, either, because 
the value $\alpha(T+\bar{T})$ is still large compared with 
$(S+\bar{S})$ and it
can not lead to positive $g^{-2}$ and $g'^{-2}$ at the same time.

Next, let us consider the $K_{5,Z \bar Z}$ dependence varying it 
for $\cos \theta =0$.
We fix $\phi$, e.g. $\tan \phi=1$.
On top of that, the gauge coupling constant of 
the observable sector $g_{GUT}$ must satisfy $2g^{-2}_{GUT} \simeq 4$.
Thus, if the field $\Phi$ belongs to the observable sector, we take 
\begin{eqnarray}
(S+\bar{S})+\alpha(T+\bar{T})=4.
\label{inf-ob}
\end{eqnarray}
On the other hand, 
if the field $\Phi$ belongs to the hidden sector, we take 
\begin{eqnarray}
(S+\bar{S})-{3 \over 2}\alpha(T+\bar{T})=4.
\label{inf-hid}
\end{eqnarray}
In the former case (\ref{inf-ob}) the positivity of 
$g^{-2}$ and $g'^{-2}$ requires $\alpha(T+\bar{T}) < 8/5$, 
while in the latter case (\ref{inf-hid}) 
$\alpha(T+\bar{T}) > -8/5$ is required.
Fig.1 shows the solution of $m_\Phi^2=0$ leading to such values of 
$\tau = \alpha(T+\bar{T})$.
The lower and upper lines correspond to the cases (\ref{inf-ob}) and 
(\ref{inf-hid}), respectively.
In the case that $\Phi$ belongs to the observable sector, 
$1/K_{5,Z \bar Z}>3.4$ is required for $\alpha(T+\bar{T}) < 8/5$.
In the case that $\Phi$ belongs to the hidden sector, 
the same region $1/K_{5,Z \bar Z}>3.4$  leads to positive values of 
$\alpha(T+\bar{T})$, that is, $\alpha(T+\bar{T}) > -8/5$ 
is satisfied.
In this region, we have the realistic solutions of $m_\Phi^2=0$ 
for $\cos \theta =0$ and $\tan \phi =1$ 
when the field $\Phi$ belongs to the observable or hidden sector.
\begin{center}
\setlength{\unitlength}{0.240900pt}
\ifx\plotpoint\undefined\newsavebox{\plotpoint}\fi
\sbox{\plotpoint}{\rule[-0.200pt]{0.400pt}{0.400pt}}%
\begin{picture}(1500,900)(0,0)
\font\gnuplot=cmr10 at 10pt
\gnuplot
\sbox{\plotpoint}{\rule[-0.200pt]{0.400pt}{0.400pt}}%
\put(161.0,123.0){\rule[-0.200pt]{4.818pt}{0.400pt}}
\put(141,123){\makebox(0,0)[r]{0}}
\put(1419.0,123.0){\rule[-0.200pt]{4.818pt}{0.400pt}}
\put(161.0,215.0){\rule[-0.200pt]{4.818pt}{0.400pt}}
\put(141,215){\makebox(0,0)[r]{0.5}}
\put(1419.0,215.0){\rule[-0.200pt]{4.818pt}{0.400pt}}
\put(161.0,307.0){\rule[-0.200pt]{4.818pt}{0.400pt}}
\put(141,307){\makebox(0,0)[r]{1}}
\put(1419.0,307.0){\rule[-0.200pt]{4.818pt}{0.400pt}}
\put(161.0,399.0){\rule[-0.200pt]{4.818pt}{0.400pt}}
\put(141,399){\makebox(0,0)[r]{1.5}}
\put(1419.0,399.0){\rule[-0.200pt]{4.818pt}{0.400pt}}
\put(161.0,492.0){\rule[-0.200pt]{4.818pt}{0.400pt}}
\put(141,492){\makebox(0,0)[r]{2}}
\put(1419.0,492.0){\rule[-0.200pt]{4.818pt}{0.400pt}}
\put(161.0,584.0){\rule[-0.200pt]{4.818pt}{0.400pt}}
\put(141,584){\makebox(0,0)[r]{2.5}}
\put(1419.0,584.0){\rule[-0.200pt]{4.818pt}{0.400pt}}
\put(161.0,676.0){\rule[-0.200pt]{4.818pt}{0.400pt}}
\put(141,676){\makebox(0,0)[r]{3}}
\put(1419.0,676.0){\rule[-0.200pt]{4.818pt}{0.400pt}}
\put(161.0,768.0){\rule[-0.200pt]{4.818pt}{0.400pt}}
\put(141,768){\makebox(0,0)[r]{3.5}}
\put(1419.0,768.0){\rule[-0.200pt]{4.818pt}{0.400pt}}
\put(161.0,860.0){\rule[-0.200pt]{4.818pt}{0.400pt}}
\put(141,860){\makebox(0,0)[r]{4}}
\put(1419.0,860.0){\rule[-0.200pt]{4.818pt}{0.400pt}}
\put(161.0,123.0){\rule[-0.200pt]{0.400pt}{4.818pt}}
\put(161,82){\makebox(0,0){0}}
\put(161.0,840.0){\rule[-0.200pt]{0.400pt}{4.818pt}}
\put(417.0,123.0){\rule[-0.200pt]{0.400pt}{4.818pt}}
\put(417,82){\makebox(0,0){2}}
\put(417.0,840.0){\rule[-0.200pt]{0.400pt}{4.818pt}}
\put(672.0,123.0){\rule[-0.200pt]{0.400pt}{4.818pt}}
\put(672,82){\makebox(0,0){4}}
\put(672.0,840.0){\rule[-0.200pt]{0.400pt}{4.818pt}}
\put(928.0,123.0){\rule[-0.200pt]{0.400pt}{4.818pt}}
\put(928,82){\makebox(0,0){6}}
\put(928.0,840.0){\rule[-0.200pt]{0.400pt}{4.818pt}}
\put(1183.0,123.0){\rule[-0.200pt]{0.400pt}{4.818pt}}
\put(1183,82){\makebox(0,0){8}}
\put(1183.0,840.0){\rule[-0.200pt]{0.400pt}{4.818pt}}
\put(1439.0,123.0){\rule[-0.200pt]{0.400pt}{4.818pt}}
\put(1439,82){\makebox(0,0){10}}
\put(1439.0,840.0){\rule[-0.200pt]{0.400pt}{4.818pt}}
\put(161.0,123.0){\rule[-0.200pt]{307.870pt}{0.400pt}}
\put(1439.0,123.0){\rule[-0.200pt]{0.400pt}{177.543pt}}
\put(161.0,860.0){\rule[-0.200pt]{307.870pt}{0.400pt}}
\put(40,491){\makebox(0,0){$\tau$}}
\put(800,21){\makebox(0,0){$1/K_{5,Z \bar Z}$}}
\put(1311,215){\makebox(0,0)[r]{observable}}
\put(1311,584){\makebox(0,0)[r]{hidden}}
\put(161.0,123.0){\rule[-0.200pt]{0.400pt}{177.543pt}}
\sbox{\plotpoint}{\rule[-0.400pt]{0.800pt}{0.800pt}}%
\put(264,750){\usebox{\plotpoint}}
\multiput(265.41,737.42)(0.509,-1.857){19}{\rule{0.123pt}{3.031pt}}
\multiput(262.34,743.71)(13.000,-39.709){2}{\rule{0.800pt}{1.515pt}}
\multiput(278.41,695.25)(0.509,-1.236){19}{\rule{0.123pt}{2.108pt}}
\multiput(275.34,699.63)(13.000,-26.625){2}{\rule{0.800pt}{1.054pt}}
\multiput(291.41,665.78)(0.509,-0.988){19}{\rule{0.123pt}{1.738pt}}
\multiput(288.34,669.39)(13.000,-21.392){2}{\rule{0.800pt}{0.869pt}}
\multiput(304.41,641.81)(0.509,-0.823){19}{\rule{0.123pt}{1.492pt}}
\multiput(301.34,644.90)(13.000,-17.903){2}{\rule{0.800pt}{0.746pt}}
\multiput(317.41,621.32)(0.509,-0.740){19}{\rule{0.123pt}{1.369pt}}
\multiput(314.34,624.16)(13.000,-16.158){2}{\rule{0.800pt}{0.685pt}}
\multiput(330.41,602.83)(0.509,-0.657){19}{\rule{0.123pt}{1.246pt}}
\multiput(327.34,605.41)(13.000,-14.414){2}{\rule{0.800pt}{0.623pt}}
\multiput(343.41,586.34)(0.509,-0.574){19}{\rule{0.123pt}{1.123pt}}
\multiput(340.34,588.67)(13.000,-12.669){2}{\rule{0.800pt}{0.562pt}}
\multiput(356.41,571.59)(0.509,-0.533){19}{\rule{0.123pt}{1.062pt}}
\multiput(353.34,573.80)(13.000,-11.797){2}{\rule{0.800pt}{0.531pt}}
\multiput(369.41,557.57)(0.511,-0.536){17}{\rule{0.123pt}{1.067pt}}
\multiput(366.34,559.79)(12.000,-10.786){2}{\rule{0.800pt}{0.533pt}}
\multiput(380.00,547.08)(0.536,-0.511){17}{\rule{1.067pt}{0.123pt}}
\multiput(380.00,547.34)(10.786,-12.000){2}{\rule{0.533pt}{0.800pt}}
\multiput(393.00,535.08)(0.589,-0.512){15}{\rule{1.145pt}{0.123pt}}
\multiput(393.00,535.34)(10.623,-11.000){2}{\rule{0.573pt}{0.800pt}}
\multiput(406.00,524.08)(0.589,-0.512){15}{\rule{1.145pt}{0.123pt}}
\multiput(406.00,524.34)(10.623,-11.000){2}{\rule{0.573pt}{0.800pt}}
\multiput(419.00,513.08)(0.737,-0.516){11}{\rule{1.356pt}{0.124pt}}
\multiput(419.00,513.34)(10.186,-9.000){2}{\rule{0.678pt}{0.800pt}}
\multiput(432.00,504.08)(0.654,-0.514){13}{\rule{1.240pt}{0.124pt}}
\multiput(432.00,504.34)(10.426,-10.000){2}{\rule{0.620pt}{0.800pt}}
\multiput(445.00,494.08)(0.847,-0.520){9}{\rule{1.500pt}{0.125pt}}
\multiput(445.00,494.34)(9.887,-8.000){2}{\rule{0.750pt}{0.800pt}}
\multiput(458.00,486.08)(0.737,-0.516){11}{\rule{1.356pt}{0.124pt}}
\multiput(458.00,486.34)(10.186,-9.000){2}{\rule{0.678pt}{0.800pt}}
\multiput(471.00,477.08)(1.000,-0.526){7}{\rule{1.686pt}{0.127pt}}
\multiput(471.00,477.34)(9.501,-7.000){2}{\rule{0.843pt}{0.800pt}}
\multiput(484.00,470.08)(0.847,-0.520){9}{\rule{1.500pt}{0.125pt}}
\multiput(484.00,470.34)(9.887,-8.000){2}{\rule{0.750pt}{0.800pt}}
\multiput(497.00,462.08)(1.000,-0.526){7}{\rule{1.686pt}{0.127pt}}
\multiput(497.00,462.34)(9.501,-7.000){2}{\rule{0.843pt}{0.800pt}}
\multiput(510.00,455.08)(0.913,-0.526){7}{\rule{1.571pt}{0.127pt}}
\multiput(510.00,455.34)(8.738,-7.000){2}{\rule{0.786pt}{0.800pt}}
\multiput(522.00,448.07)(1.244,-0.536){5}{\rule{1.933pt}{0.129pt}}
\multiput(522.00,448.34)(8.987,-6.000){2}{\rule{0.967pt}{0.800pt}}
\multiput(535.00,442.07)(1.244,-0.536){5}{\rule{1.933pt}{0.129pt}}
\multiput(535.00,442.34)(8.987,-6.000){2}{\rule{0.967pt}{0.800pt}}
\multiput(548.00,436.07)(1.244,-0.536){5}{\rule{1.933pt}{0.129pt}}
\multiput(548.00,436.34)(8.987,-6.000){2}{\rule{0.967pt}{0.800pt}}
\multiput(561.00,430.07)(1.244,-0.536){5}{\rule{1.933pt}{0.129pt}}
\multiput(561.00,430.34)(8.987,-6.000){2}{\rule{0.967pt}{0.800pt}}
\multiput(574.00,424.06)(1.768,-0.560){3}{\rule{2.280pt}{0.135pt}}
\multiput(574.00,424.34)(8.268,-5.000){2}{\rule{1.140pt}{0.800pt}}
\multiput(587.00,419.06)(1.768,-0.560){3}{\rule{2.280pt}{0.135pt}}
\multiput(587.00,419.34)(8.268,-5.000){2}{\rule{1.140pt}{0.800pt}}
\multiput(600.00,414.06)(1.768,-0.560){3}{\rule{2.280pt}{0.135pt}}
\multiput(600.00,414.34)(8.268,-5.000){2}{\rule{1.140pt}{0.800pt}}
\multiput(613.00,409.06)(1.768,-0.560){3}{\rule{2.280pt}{0.135pt}}
\multiput(613.00,409.34)(8.268,-5.000){2}{\rule{1.140pt}{0.800pt}}
\multiput(626.00,404.06)(1.768,-0.560){3}{\rule{2.280pt}{0.135pt}}
\multiput(626.00,404.34)(8.268,-5.000){2}{\rule{1.140pt}{0.800pt}}
\put(639,397.34){\rule{2.800pt}{0.800pt}}
\multiput(639.00,399.34)(7.188,-4.000){2}{\rule{1.400pt}{0.800pt}}
\multiput(652.00,395.06)(1.600,-0.560){3}{\rule{2.120pt}{0.135pt}}
\multiput(652.00,395.34)(7.600,-5.000){2}{\rule{1.060pt}{0.800pt}}
\put(664,388.34){\rule{2.800pt}{0.800pt}}
\multiput(664.00,390.34)(7.188,-4.000){2}{\rule{1.400pt}{0.800pt}}
\put(677,384.34){\rule{2.800pt}{0.800pt}}
\multiput(677.00,386.34)(7.188,-4.000){2}{\rule{1.400pt}{0.800pt}}
\put(690,380.34){\rule{2.800pt}{0.800pt}}
\multiput(690.00,382.34)(7.188,-4.000){2}{\rule{1.400pt}{0.800pt}}
\put(703,376.84){\rule{3.132pt}{0.800pt}}
\multiput(703.00,378.34)(6.500,-3.000){2}{\rule{1.566pt}{0.800pt}}
\put(716,373.34){\rule{2.800pt}{0.800pt}}
\multiput(716.00,375.34)(7.188,-4.000){2}{\rule{1.400pt}{0.800pt}}
\put(729,369.34){\rule{2.800pt}{0.800pt}}
\multiput(729.00,371.34)(7.188,-4.000){2}{\rule{1.400pt}{0.800pt}}
\put(742,365.84){\rule{3.132pt}{0.800pt}}
\multiput(742.00,367.34)(6.500,-3.000){2}{\rule{1.566pt}{0.800pt}}
\put(755,362.84){\rule{3.132pt}{0.800pt}}
\multiput(755.00,364.34)(6.500,-3.000){2}{\rule{1.566pt}{0.800pt}}
\put(768,359.34){\rule{2.800pt}{0.800pt}}
\multiput(768.00,361.34)(7.188,-4.000){2}{\rule{1.400pt}{0.800pt}}
\put(781,355.84){\rule{3.132pt}{0.800pt}}
\multiput(781.00,357.34)(6.500,-3.000){2}{\rule{1.566pt}{0.800pt}}
\put(794,352.84){\rule{2.891pt}{0.800pt}}
\multiput(794.00,354.34)(6.000,-3.000){2}{\rule{1.445pt}{0.800pt}}
\put(806,349.84){\rule{3.132pt}{0.800pt}}
\multiput(806.00,351.34)(6.500,-3.000){2}{\rule{1.566pt}{0.800pt}}
\put(819,346.84){\rule{3.132pt}{0.800pt}}
\multiput(819.00,348.34)(6.500,-3.000){2}{\rule{1.566pt}{0.800pt}}
\put(832,344.34){\rule{3.132pt}{0.800pt}}
\multiput(832.00,345.34)(6.500,-2.000){2}{\rule{1.566pt}{0.800pt}}
\put(845,341.84){\rule{3.132pt}{0.800pt}}
\multiput(845.00,343.34)(6.500,-3.000){2}{\rule{1.566pt}{0.800pt}}
\put(858,338.84){\rule{3.132pt}{0.800pt}}
\multiput(858.00,340.34)(6.500,-3.000){2}{\rule{1.566pt}{0.800pt}}
\put(871,336.34){\rule{3.132pt}{0.800pt}}
\multiput(871.00,337.34)(6.500,-2.000){2}{\rule{1.566pt}{0.800pt}}
\put(884,333.84){\rule{3.132pt}{0.800pt}}
\multiput(884.00,335.34)(6.500,-3.000){2}{\rule{1.566pt}{0.800pt}}
\put(897,331.34){\rule{3.132pt}{0.800pt}}
\multiput(897.00,332.34)(6.500,-2.000){2}{\rule{1.566pt}{0.800pt}}
\put(910,328.84){\rule{3.132pt}{0.800pt}}
\multiput(910.00,330.34)(6.500,-3.000){2}{\rule{1.566pt}{0.800pt}}
\put(923,326.34){\rule{3.132pt}{0.800pt}}
\multiput(923.00,327.34)(6.500,-2.000){2}{\rule{1.566pt}{0.800pt}}
\put(936,324.34){\rule{2.891pt}{0.800pt}}
\multiput(936.00,325.34)(6.000,-2.000){2}{\rule{1.445pt}{0.800pt}}
\put(948,321.84){\rule{3.132pt}{0.800pt}}
\multiput(948.00,323.34)(6.500,-3.000){2}{\rule{1.566pt}{0.800pt}}
\put(961,319.34){\rule{3.132pt}{0.800pt}}
\multiput(961.00,320.34)(6.500,-2.000){2}{\rule{1.566pt}{0.800pt}}
\put(974,317.34){\rule{3.132pt}{0.800pt}}
\multiput(974.00,318.34)(6.500,-2.000){2}{\rule{1.566pt}{0.800pt}}
\put(987,315.34){\rule{3.132pt}{0.800pt}}
\multiput(987.00,316.34)(6.500,-2.000){2}{\rule{1.566pt}{0.800pt}}
\put(1000,313.34){\rule{3.132pt}{0.800pt}}
\multiput(1000.00,314.34)(6.500,-2.000){2}{\rule{1.566pt}{0.800pt}}
\put(1013,311.34){\rule{3.132pt}{0.800pt}}
\multiput(1013.00,312.34)(6.500,-2.000){2}{\rule{1.566pt}{0.800pt}}
\put(1026,309.34){\rule{3.132pt}{0.800pt}}
\multiput(1026.00,310.34)(6.500,-2.000){2}{\rule{1.566pt}{0.800pt}}
\put(1039,307.34){\rule{3.132pt}{0.800pt}}
\multiput(1039.00,308.34)(6.500,-2.000){2}{\rule{1.566pt}{0.800pt}}
\put(1052,305.34){\rule{3.132pt}{0.800pt}}
\multiput(1052.00,306.34)(6.500,-2.000){2}{\rule{1.566pt}{0.800pt}}
\put(1065,303.34){\rule{3.132pt}{0.800pt}}
\multiput(1065.00,304.34)(6.500,-2.000){2}{\rule{1.566pt}{0.800pt}}
\put(1078,301.84){\rule{2.891pt}{0.800pt}}
\multiput(1078.00,302.34)(6.000,-1.000){2}{\rule{1.445pt}{0.800pt}}
\put(1090,300.34){\rule{3.132pt}{0.800pt}}
\multiput(1090.00,301.34)(6.500,-2.000){2}{\rule{1.566pt}{0.800pt}}
\put(1103,298.34){\rule{3.132pt}{0.800pt}}
\multiput(1103.00,299.34)(6.500,-2.000){2}{\rule{1.566pt}{0.800pt}}
\put(1116,296.84){\rule{3.132pt}{0.800pt}}
\multiput(1116.00,297.34)(6.500,-1.000){2}{\rule{1.566pt}{0.800pt}}
\put(1129,295.34){\rule{3.132pt}{0.800pt}}
\multiput(1129.00,296.34)(6.500,-2.000){2}{\rule{1.566pt}{0.800pt}}
\put(1142,293.34){\rule{3.132pt}{0.800pt}}
\multiput(1142.00,294.34)(6.500,-2.000){2}{\rule{1.566pt}{0.800pt}}
\put(1155,291.84){\rule{3.132pt}{0.800pt}}
\multiput(1155.00,292.34)(6.500,-1.000){2}{\rule{1.566pt}{0.800pt}}
\put(1168,290.34){\rule{3.132pt}{0.800pt}}
\multiput(1168.00,291.34)(6.500,-2.000){2}{\rule{1.566pt}{0.800pt}}
\put(1181,288.84){\rule{3.132pt}{0.800pt}}
\multiput(1181.00,289.34)(6.500,-1.000){2}{\rule{1.566pt}{0.800pt}}
\put(1194,287.34){\rule{3.132pt}{0.800pt}}
\multiput(1194.00,288.34)(6.500,-2.000){2}{\rule{1.566pt}{0.800pt}}
\put(1207,285.84){\rule{3.132pt}{0.800pt}}
\multiput(1207.00,286.34)(6.500,-1.000){2}{\rule{1.566pt}{0.800pt}}
\put(1220,284.34){\rule{2.891pt}{0.800pt}}
\multiput(1220.00,285.34)(6.000,-2.000){2}{\rule{1.445pt}{0.800pt}}
\put(1232,282.84){\rule{3.132pt}{0.800pt}}
\multiput(1232.00,283.34)(6.500,-1.000){2}{\rule{1.566pt}{0.800pt}}
\put(1245,281.34){\rule{3.132pt}{0.800pt}}
\multiput(1245.00,282.34)(6.500,-2.000){2}{\rule{1.566pt}{0.800pt}}
\put(1258,279.84){\rule{3.132pt}{0.800pt}}
\multiput(1258.00,280.34)(6.500,-1.000){2}{\rule{1.566pt}{0.800pt}}
\put(1271,278.84){\rule{3.132pt}{0.800pt}}
\multiput(1271.00,279.34)(6.500,-1.000){2}{\rule{1.566pt}{0.800pt}}
\put(1284,277.34){\rule{3.132pt}{0.800pt}}
\multiput(1284.00,278.34)(6.500,-2.000){2}{\rule{1.566pt}{0.800pt}}
\put(1297,275.84){\rule{3.132pt}{0.800pt}}
\multiput(1297.00,276.34)(6.500,-1.000){2}{\rule{1.566pt}{0.800pt}}
\put(1310,274.84){\rule{3.132pt}{0.800pt}}
\multiput(1310.00,275.34)(6.500,-1.000){2}{\rule{1.566pt}{0.800pt}}
\put(1323,273.84){\rule{3.132pt}{0.800pt}}
\multiput(1323.00,274.34)(6.500,-1.000){2}{\rule{1.566pt}{0.800pt}}
\put(1336,272.34){\rule{3.132pt}{0.800pt}}
\multiput(1336.00,273.34)(6.500,-2.000){2}{\rule{1.566pt}{0.800pt}}
\put(1349,270.84){\rule{3.132pt}{0.800pt}}
\multiput(1349.00,271.34)(6.500,-1.000){2}{\rule{1.566pt}{0.800pt}}
\put(1362,269.84){\rule{2.891pt}{0.800pt}}
\multiput(1362.00,270.34)(6.000,-1.000){2}{\rule{1.445pt}{0.800pt}}
\put(1374,268.84){\rule{3.132pt}{0.800pt}}
\multiput(1374.00,269.34)(6.500,-1.000){2}{\rule{1.566pt}{0.800pt}}
\put(1387,267.84){\rule{3.132pt}{0.800pt}}
\multiput(1387.00,268.34)(6.500,-1.000){2}{\rule{1.566pt}{0.800pt}}
\put(1400,266.84){\rule{3.132pt}{0.800pt}}
\multiput(1400.00,267.34)(6.500,-1.000){2}{\rule{1.566pt}{0.800pt}}
\put(1413,265.84){\rule{3.132pt}{0.800pt}}
\multiput(1413.00,266.34)(6.500,-1.000){2}{\rule{1.566pt}{0.800pt}}
\put(1426,264.34){\rule{3.132pt}{0.800pt}}
\multiput(1426.00,265.34)(6.500,-2.000){2}{\rule{1.566pt}{0.800pt}}
\multiput(900.40,852.23)(0.512,-1.088){15}{\rule{0.123pt}{1.873pt}}
\multiput(897.34,856.11)(11.000,-19.113){2}{\rule{0.800pt}{0.936pt}}
\multiput(911.41,829.02)(0.509,-1.112){19}{\rule{0.123pt}{1.923pt}}
\multiput(908.34,833.01)(13.000,-24.009){2}{\rule{0.800pt}{0.962pt}}
\multiput(924.41,801.78)(0.509,-0.988){19}{\rule{0.123pt}{1.738pt}}
\multiput(921.34,805.39)(13.000,-21.392){2}{\rule{0.800pt}{0.869pt}}
\multiput(937.41,776.80)(0.511,-0.988){17}{\rule{0.123pt}{1.733pt}}
\multiput(934.34,780.40)(12.000,-19.402){2}{\rule{0.800pt}{0.867pt}}
\multiput(949.41,754.55)(0.509,-0.864){19}{\rule{0.123pt}{1.554pt}}
\multiput(946.34,757.77)(13.000,-18.775){2}{\rule{0.800pt}{0.777pt}}
\multiput(962.41,733.06)(0.509,-0.781){19}{\rule{0.123pt}{1.431pt}}
\multiput(959.34,736.03)(13.000,-17.030){2}{\rule{0.800pt}{0.715pt}}
\multiput(975.41,713.32)(0.509,-0.740){19}{\rule{0.123pt}{1.369pt}}
\multiput(972.34,716.16)(13.000,-16.158){2}{\rule{0.800pt}{0.685pt}}
\multiput(988.41,694.57)(0.509,-0.698){19}{\rule{0.123pt}{1.308pt}}
\multiput(985.34,697.29)(13.000,-15.286){2}{\rule{0.800pt}{0.654pt}}
\multiput(1001.41,676.83)(0.509,-0.657){19}{\rule{0.123pt}{1.246pt}}
\multiput(998.34,679.41)(13.000,-14.414){2}{\rule{0.800pt}{0.623pt}}
\multiput(1014.41,660.08)(0.509,-0.616){19}{\rule{0.123pt}{1.185pt}}
\multiput(1011.34,662.54)(13.000,-13.541){2}{\rule{0.800pt}{0.592pt}}
\multiput(1027.41,644.59)(0.509,-0.533){19}{\rule{0.123pt}{1.062pt}}
\multiput(1024.34,646.80)(13.000,-11.797){2}{\rule{0.800pt}{0.531pt}}
\multiput(1040.41,630.59)(0.509,-0.533){19}{\rule{0.123pt}{1.062pt}}
\multiput(1037.34,632.80)(13.000,-11.797){2}{\rule{0.800pt}{0.531pt}}
\multiput(1053.41,616.59)(0.509,-0.533){19}{\rule{0.123pt}{1.062pt}}
\multiput(1050.34,618.80)(13.000,-11.797){2}{\rule{0.800pt}{0.531pt}}
\multiput(1065.00,605.08)(0.536,-0.511){17}{\rule{1.067pt}{0.123pt}}
\multiput(1065.00,605.34)(10.786,-12.000){2}{\rule{0.533pt}{0.800pt}}
\multiput(1078.00,593.08)(0.491,-0.511){17}{\rule{1.000pt}{0.123pt}}
\multiput(1078.00,593.34)(9.924,-12.000){2}{\rule{0.500pt}{0.800pt}}
\multiput(1090.00,581.08)(0.589,-0.512){15}{\rule{1.145pt}{0.123pt}}
\multiput(1090.00,581.34)(10.623,-11.000){2}{\rule{0.573pt}{0.800pt}}
\multiput(1103.00,570.08)(0.589,-0.512){15}{\rule{1.145pt}{0.123pt}}
\multiput(1103.00,570.34)(10.623,-11.000){2}{\rule{0.573pt}{0.800pt}}
\multiput(1116.00,559.08)(0.654,-0.514){13}{\rule{1.240pt}{0.124pt}}
\multiput(1116.00,559.34)(10.426,-10.000){2}{\rule{0.620pt}{0.800pt}}
\multiput(1129.00,549.08)(0.654,-0.514){13}{\rule{1.240pt}{0.124pt}}
\multiput(1129.00,549.34)(10.426,-10.000){2}{\rule{0.620pt}{0.800pt}}
\multiput(1142.00,539.08)(0.737,-0.516){11}{\rule{1.356pt}{0.124pt}}
\multiput(1142.00,539.34)(10.186,-9.000){2}{\rule{0.678pt}{0.800pt}}
\multiput(1155.00,530.08)(0.737,-0.516){11}{\rule{1.356pt}{0.124pt}}
\multiput(1155.00,530.34)(10.186,-9.000){2}{\rule{0.678pt}{0.800pt}}
\multiput(1168.00,521.08)(0.737,-0.516){11}{\rule{1.356pt}{0.124pt}}
\multiput(1168.00,521.34)(10.186,-9.000){2}{\rule{0.678pt}{0.800pt}}
\multiput(1181.00,512.08)(0.847,-0.520){9}{\rule{1.500pt}{0.125pt}}
\multiput(1181.00,512.34)(9.887,-8.000){2}{\rule{0.750pt}{0.800pt}}
\multiput(1194.00,504.08)(1.000,-0.526){7}{\rule{1.686pt}{0.127pt}}
\multiput(1194.00,504.34)(9.501,-7.000){2}{\rule{0.843pt}{0.800pt}}
\multiput(1207.00,497.08)(0.847,-0.520){9}{\rule{1.500pt}{0.125pt}}
\multiput(1207.00,497.34)(9.887,-8.000){2}{\rule{0.750pt}{0.800pt}}
\multiput(1220.00,489.08)(0.913,-0.526){7}{\rule{1.571pt}{0.127pt}}
\multiput(1220.00,489.34)(8.738,-7.000){2}{\rule{0.786pt}{0.800pt}}
\multiput(1232.00,482.08)(1.000,-0.526){7}{\rule{1.686pt}{0.127pt}}
\multiput(1232.00,482.34)(9.501,-7.000){2}{\rule{0.843pt}{0.800pt}}
\multiput(1245.00,475.08)(1.000,-0.526){7}{\rule{1.686pt}{0.127pt}}
\multiput(1245.00,475.34)(9.501,-7.000){2}{\rule{0.843pt}{0.800pt}}
\multiput(1258.00,468.07)(1.244,-0.536){5}{\rule{1.933pt}{0.129pt}}
\multiput(1258.00,468.34)(8.987,-6.000){2}{\rule{0.967pt}{0.800pt}}
\multiput(1271.00,462.07)(1.244,-0.536){5}{\rule{1.933pt}{0.129pt}}
\multiput(1271.00,462.34)(8.987,-6.000){2}{\rule{0.967pt}{0.800pt}}
\multiput(1284.00,456.07)(1.244,-0.536){5}{\rule{1.933pt}{0.129pt}}
\multiput(1284.00,456.34)(8.987,-6.000){2}{\rule{0.967pt}{0.800pt}}
\multiput(1297.00,450.07)(1.244,-0.536){5}{\rule{1.933pt}{0.129pt}}
\multiput(1297.00,450.34)(8.987,-6.000){2}{\rule{0.967pt}{0.800pt}}
\multiput(1310.00,444.06)(1.768,-0.560){3}{\rule{2.280pt}{0.135pt}}
\multiput(1310.00,444.34)(8.268,-5.000){2}{\rule{1.140pt}{0.800pt}}
\multiput(1323.00,439.07)(1.244,-0.536){5}{\rule{1.933pt}{0.129pt}}
\multiput(1323.00,439.34)(8.987,-6.000){2}{\rule{0.967pt}{0.800pt}}
\multiput(1336.00,433.06)(1.768,-0.560){3}{\rule{2.280pt}{0.135pt}}
\multiput(1336.00,433.34)(8.268,-5.000){2}{\rule{1.140pt}{0.800pt}}
\multiput(1349.00,428.06)(1.768,-0.560){3}{\rule{2.280pt}{0.135pt}}
\multiput(1349.00,428.34)(8.268,-5.000){2}{\rule{1.140pt}{0.800pt}}
\multiput(1362.00,423.06)(1.600,-0.560){3}{\rule{2.120pt}{0.135pt}}
\multiput(1362.00,423.34)(7.600,-5.000){2}{\rule{1.060pt}{0.800pt}}
\put(1374,416.34){\rule{2.800pt}{0.800pt}}
\multiput(1374.00,418.34)(7.188,-4.000){2}{\rule{1.400pt}{0.800pt}}
\multiput(1387.00,414.06)(1.768,-0.560){3}{\rule{2.280pt}{0.135pt}}
\multiput(1387.00,414.34)(8.268,-5.000){2}{\rule{1.140pt}{0.800pt}}
\put(1400,407.34){\rule{2.800pt}{0.800pt}}
\multiput(1400.00,409.34)(7.188,-4.000){2}{\rule{1.400pt}{0.800pt}}
\put(1413,403.34){\rule{2.800pt}{0.800pt}}
\multiput(1413.00,405.34)(7.188,-4.000){2}{\rule{1.400pt}{0.800pt}}
\put(1426,399.34){\rule{2.800pt}{0.800pt}}
\multiput(1426.00,401.34)(7.188,-4.000){2}{\rule{1.400pt}{0.800pt}}
\end{picture}

Fig.1: Solutions of $m_\Phi^2=0$ for $\cos \theta =0$.
\end{center}

Similarly, we can investigate the condition $m_\Phi^2=0$ for 
other values of $\theta$.
Let us discuss the case with $\tan \theta =1/\sqrt 3$ as 
another example.
We concentrate to the case that $\Phi$ belongs 
to the observable sector.
When we fix $K_{5,Z \bar Z}=1$, we have only the non-realistic solution, 
${(S+\bar{S})/ (\alpha(T+\bar{T})}) =0$,
i.e. $\alpha(T+\bar{T})=4$.
Now let us vary $K_{5,Z \bar Z}$ fixing $\phi$, e.g. $\tan \phi = 1$.
Fig. 2 show the realistic solution against $K_{5,Z \bar Z}$ 
for $\tan \phi =1$.
In this case, the region $1/K_{5,Z \bar Z}>5$ is favorable. 
\begin{center}
\setlength{\unitlength}{0.240900pt}
\ifx\plotpoint\undefined\newsavebox{\plotpoint}\fi
\sbox{\plotpoint}{\rule[-0.200pt]{0.400pt}{0.400pt}}%
\begin{picture}(1500,900)(0,0)
\font\gnuplot=cmr10 at 10pt
\gnuplot
\sbox{\plotpoint}{\rule[-0.200pt]{0.400pt}{0.400pt}}%
\put(161.0,123.0){\rule[-0.200pt]{4.818pt}{0.400pt}}
\put(141,123){\makebox(0,0)[r]{0}}
\put(1419.0,123.0){\rule[-0.200pt]{4.818pt}{0.400pt}}
\put(161.0,215.0){\rule[-0.200pt]{4.818pt}{0.400pt}}
\put(141,215){\makebox(0,0)[r]{0.5}}
\put(1419.0,215.0){\rule[-0.200pt]{4.818pt}{0.400pt}}
\put(161.0,307.0){\rule[-0.200pt]{4.818pt}{0.400pt}}
\put(141,307){\makebox(0,0)[r]{1}}
\put(1419.0,307.0){\rule[-0.200pt]{4.818pt}{0.400pt}}
\put(161.0,399.0){\rule[-0.200pt]{4.818pt}{0.400pt}}
\put(141,399){\makebox(0,0)[r]{1.5}}
\put(1419.0,399.0){\rule[-0.200pt]{4.818pt}{0.400pt}}
\put(161.0,492.0){\rule[-0.200pt]{4.818pt}{0.400pt}}
\put(141,492){\makebox(0,0)[r]{2}}
\put(1419.0,492.0){\rule[-0.200pt]{4.818pt}{0.400pt}}
\put(161.0,584.0){\rule[-0.200pt]{4.818pt}{0.400pt}}
\put(141,584){\makebox(0,0)[r]{2.5}}
\put(1419.0,584.0){\rule[-0.200pt]{4.818pt}{0.400pt}}
\put(161.0,676.0){\rule[-0.200pt]{4.818pt}{0.400pt}}
\put(141,676){\makebox(0,0)[r]{3}}
\put(1419.0,676.0){\rule[-0.200pt]{4.818pt}{0.400pt}}
\put(161.0,768.0){\rule[-0.200pt]{4.818pt}{0.400pt}}
\put(141,768){\makebox(0,0)[r]{3.5}}
\put(1419.0,768.0){\rule[-0.200pt]{4.818pt}{0.400pt}}
\put(161.0,860.0){\rule[-0.200pt]{4.818pt}{0.400pt}}
\put(141,860){\makebox(0,0)[r]{4}}
\put(1419.0,860.0){\rule[-0.200pt]{4.818pt}{0.400pt}}
\put(161.0,123.0){\rule[-0.200pt]{0.400pt}{4.818pt}}
\put(161,82){\makebox(0,0){0}}
\put(161.0,840.0){\rule[-0.200pt]{0.400pt}{4.818pt}}
\put(417.0,123.0){\rule[-0.200pt]{0.400pt}{4.818pt}}
\put(417,82){\makebox(0,0){2}}
\put(417.0,840.0){\rule[-0.200pt]{0.400pt}{4.818pt}}
\put(672.0,123.0){\rule[-0.200pt]{0.400pt}{4.818pt}}
\put(672,82){\makebox(0,0){4}}
\put(672.0,840.0){\rule[-0.200pt]{0.400pt}{4.818pt}}
\put(928.0,123.0){\rule[-0.200pt]{0.400pt}{4.818pt}}
\put(928,82){\makebox(0,0){6}}
\put(928.0,840.0){\rule[-0.200pt]{0.400pt}{4.818pt}}
\put(1183.0,123.0){\rule[-0.200pt]{0.400pt}{4.818pt}}
\put(1183,82){\makebox(0,0){8}}
\put(1183.0,840.0){\rule[-0.200pt]{0.400pt}{4.818pt}}
\put(1439.0,123.0){\rule[-0.200pt]{0.400pt}{4.818pt}}
\put(1439,82){\makebox(0,0){10}}
\put(1439.0,840.0){\rule[-0.200pt]{0.400pt}{4.818pt}}
\put(161.0,123.0){\rule[-0.200pt]{307.870pt}{0.400pt}}
\put(1439.0,123.0){\rule[-0.200pt]{0.400pt}{177.543pt}}
\put(161.0,860.0){\rule[-0.200pt]{307.870pt}{0.400pt}}
\put(40,491){\makebox(0,0){$\tau$}}
\put(800,21){\makebox(0,0){$1/K_{5,Z \bar Z}$}}
\put(161.0,123.0){\rule[-0.200pt]{0.400pt}{177.543pt}}
\sbox{\plotpoint}{\rule[-0.400pt]{0.800pt}{0.800pt}}%
\put(380,852){\usebox{\plotpoint}}
\multiput(381.41,836.35)(0.509,-2.353){19}{\rule{0.123pt}{3.769pt}}
\multiput(378.34,844.18)(13.000,-50.177){2}{\rule{0.800pt}{1.885pt}}
\multiput(394.41,785.00)(0.509,-1.278){19}{\rule{0.123pt}{2.169pt}}
\multiput(391.34,789.50)(13.000,-27.498){2}{\rule{0.800pt}{1.085pt}}
\multiput(407.41,754.53)(0.509,-1.029){19}{\rule{0.123pt}{1.800pt}}
\multiput(404.34,758.26)(13.000,-22.264){2}{\rule{0.800pt}{0.900pt}}
\multiput(420.41,729.29)(0.509,-0.905){19}{\rule{0.123pt}{1.615pt}}
\multiput(417.34,732.65)(13.000,-19.647){2}{\rule{0.800pt}{0.808pt}}
\multiput(433.41,707.06)(0.509,-0.781){19}{\rule{0.123pt}{1.431pt}}
\multiput(430.34,710.03)(13.000,-17.030){2}{\rule{0.800pt}{0.715pt}}
\multiput(446.41,687.32)(0.509,-0.740){19}{\rule{0.123pt}{1.369pt}}
\multiput(443.34,690.16)(13.000,-16.158){2}{\rule{0.800pt}{0.685pt}}
\multiput(459.41,668.83)(0.509,-0.657){19}{\rule{0.123pt}{1.246pt}}
\multiput(456.34,671.41)(13.000,-14.414){2}{\rule{0.800pt}{0.623pt}}
\multiput(472.41,652.34)(0.509,-0.574){19}{\rule{0.123pt}{1.123pt}}
\multiput(469.34,654.67)(13.000,-12.669){2}{\rule{0.800pt}{0.562pt}}
\multiput(485.41,637.34)(0.509,-0.574){19}{\rule{0.123pt}{1.123pt}}
\multiput(482.34,639.67)(13.000,-12.669){2}{\rule{0.800pt}{0.562pt}}
\multiput(498.41,622.59)(0.509,-0.533){19}{\rule{0.123pt}{1.062pt}}
\multiput(495.34,624.80)(13.000,-11.797){2}{\rule{0.800pt}{0.531pt}}
\multiput(511.41,608.57)(0.511,-0.536){17}{\rule{0.123pt}{1.067pt}}
\multiput(508.34,610.79)(12.000,-10.786){2}{\rule{0.800pt}{0.533pt}}
\multiput(522.00,598.08)(0.536,-0.511){17}{\rule{1.067pt}{0.123pt}}
\multiput(522.00,598.34)(10.786,-12.000){2}{\rule{0.533pt}{0.800pt}}
\multiput(535.00,586.08)(0.589,-0.512){15}{\rule{1.145pt}{0.123pt}}
\multiput(535.00,586.34)(10.623,-11.000){2}{\rule{0.573pt}{0.800pt}}
\multiput(548.00,575.08)(0.589,-0.512){15}{\rule{1.145pt}{0.123pt}}
\multiput(548.00,575.34)(10.623,-11.000){2}{\rule{0.573pt}{0.800pt}}
\multiput(561.00,564.08)(0.589,-0.512){15}{\rule{1.145pt}{0.123pt}}
\multiput(561.00,564.34)(10.623,-11.000){2}{\rule{0.573pt}{0.800pt}}
\multiput(574.00,553.08)(0.654,-0.514){13}{\rule{1.240pt}{0.124pt}}
\multiput(574.00,553.34)(10.426,-10.000){2}{\rule{0.620pt}{0.800pt}}
\multiput(587.00,543.08)(0.737,-0.516){11}{\rule{1.356pt}{0.124pt}}
\multiput(587.00,543.34)(10.186,-9.000){2}{\rule{0.678pt}{0.800pt}}
\multiput(600.00,534.08)(0.737,-0.516){11}{\rule{1.356pt}{0.124pt}}
\multiput(600.00,534.34)(10.186,-9.000){2}{\rule{0.678pt}{0.800pt}}
\multiput(613.00,525.08)(0.847,-0.520){9}{\rule{1.500pt}{0.125pt}}
\multiput(613.00,525.34)(9.887,-8.000){2}{\rule{0.750pt}{0.800pt}}
\multiput(626.00,517.08)(0.737,-0.516){11}{\rule{1.356pt}{0.124pt}}
\multiput(626.00,517.34)(10.186,-9.000){2}{\rule{0.678pt}{0.800pt}}
\multiput(639.00,508.08)(1.000,-0.526){7}{\rule{1.686pt}{0.127pt}}
\multiput(639.00,508.34)(9.501,-7.000){2}{\rule{0.843pt}{0.800pt}}
\multiput(652.00,501.08)(0.774,-0.520){9}{\rule{1.400pt}{0.125pt}}
\multiput(652.00,501.34)(9.094,-8.000){2}{\rule{0.700pt}{0.800pt}}
\multiput(664.00,493.08)(1.000,-0.526){7}{\rule{1.686pt}{0.127pt}}
\multiput(664.00,493.34)(9.501,-7.000){2}{\rule{0.843pt}{0.800pt}}
\multiput(677.00,486.08)(1.000,-0.526){7}{\rule{1.686pt}{0.127pt}}
\multiput(677.00,486.34)(9.501,-7.000){2}{\rule{0.843pt}{0.800pt}}
\multiput(690.00,479.08)(1.000,-0.526){7}{\rule{1.686pt}{0.127pt}}
\multiput(690.00,479.34)(9.501,-7.000){2}{\rule{0.843pt}{0.800pt}}
\multiput(703.00,472.07)(1.244,-0.536){5}{\rule{1.933pt}{0.129pt}}
\multiput(703.00,472.34)(8.987,-6.000){2}{\rule{0.967pt}{0.800pt}}
\multiput(716.00,466.07)(1.244,-0.536){5}{\rule{1.933pt}{0.129pt}}
\multiput(716.00,466.34)(8.987,-6.000){2}{\rule{0.967pt}{0.800pt}}
\multiput(729.00,460.07)(1.244,-0.536){5}{\rule{1.933pt}{0.129pt}}
\multiput(729.00,460.34)(8.987,-6.000){2}{\rule{0.967pt}{0.800pt}}
\multiput(742.00,454.07)(1.244,-0.536){5}{\rule{1.933pt}{0.129pt}}
\multiput(742.00,454.34)(8.987,-6.000){2}{\rule{0.967pt}{0.800pt}}
\multiput(755.00,448.07)(1.244,-0.536){5}{\rule{1.933pt}{0.129pt}}
\multiput(755.00,448.34)(8.987,-6.000){2}{\rule{0.967pt}{0.800pt}}
\multiput(768.00,442.06)(1.768,-0.560){3}{\rule{2.280pt}{0.135pt}}
\multiput(768.00,442.34)(8.268,-5.000){2}{\rule{1.140pt}{0.800pt}}
\multiput(781.00,437.06)(1.768,-0.560){3}{\rule{2.280pt}{0.135pt}}
\multiput(781.00,437.34)(8.268,-5.000){2}{\rule{1.140pt}{0.800pt}}
\multiput(794.00,432.06)(1.600,-0.560){3}{\rule{2.120pt}{0.135pt}}
\multiput(794.00,432.34)(7.600,-5.000){2}{\rule{1.060pt}{0.800pt}}
\multiput(806.00,427.06)(1.768,-0.560){3}{\rule{2.280pt}{0.135pt}}
\multiput(806.00,427.34)(8.268,-5.000){2}{\rule{1.140pt}{0.800pt}}
\multiput(819.00,422.06)(1.768,-0.560){3}{\rule{2.280pt}{0.135pt}}
\multiput(819.00,422.34)(8.268,-5.000){2}{\rule{1.140pt}{0.800pt}}
\put(832,415.34){\rule{2.800pt}{0.800pt}}
\multiput(832.00,417.34)(7.188,-4.000){2}{\rule{1.400pt}{0.800pt}}
\multiput(845.00,413.06)(1.768,-0.560){3}{\rule{2.280pt}{0.135pt}}
\multiput(845.00,413.34)(8.268,-5.000){2}{\rule{1.140pt}{0.800pt}}
\put(858,406.34){\rule{2.800pt}{0.800pt}}
\multiput(858.00,408.34)(7.188,-4.000){2}{\rule{1.400pt}{0.800pt}}
\put(871,402.34){\rule{2.800pt}{0.800pt}}
\multiput(871.00,404.34)(7.188,-4.000){2}{\rule{1.400pt}{0.800pt}}
\put(884,398.34){\rule{2.800pt}{0.800pt}}
\multiput(884.00,400.34)(7.188,-4.000){2}{\rule{1.400pt}{0.800pt}}
\put(897,394.34){\rule{2.800pt}{0.800pt}}
\multiput(897.00,396.34)(7.188,-4.000){2}{\rule{1.400pt}{0.800pt}}
\put(910,390.34){\rule{2.800pt}{0.800pt}}
\multiput(910.00,392.34)(7.188,-4.000){2}{\rule{1.400pt}{0.800pt}}
\put(923,386.34){\rule{2.800pt}{0.800pt}}
\multiput(923.00,388.34)(7.188,-4.000){2}{\rule{1.400pt}{0.800pt}}
\put(936,382.84){\rule{2.891pt}{0.800pt}}
\multiput(936.00,384.34)(6.000,-3.000){2}{\rule{1.445pt}{0.800pt}}
\put(948,379.34){\rule{2.800pt}{0.800pt}}
\multiput(948.00,381.34)(7.188,-4.000){2}{\rule{1.400pt}{0.800pt}}
\put(961,375.84){\rule{3.132pt}{0.800pt}}
\multiput(961.00,377.34)(6.500,-3.000){2}{\rule{1.566pt}{0.800pt}}
\put(974,372.84){\rule{3.132pt}{0.800pt}}
\multiput(974.00,374.34)(6.500,-3.000){2}{\rule{1.566pt}{0.800pt}}
\put(987,369.34){\rule{2.800pt}{0.800pt}}
\multiput(987.00,371.34)(7.188,-4.000){2}{\rule{1.400pt}{0.800pt}}
\put(1000,365.84){\rule{3.132pt}{0.800pt}}
\multiput(1000.00,367.34)(6.500,-3.000){2}{\rule{1.566pt}{0.800pt}}
\put(1013,362.84){\rule{3.132pt}{0.800pt}}
\multiput(1013.00,364.34)(6.500,-3.000){2}{\rule{1.566pt}{0.800pt}}
\put(1026,359.84){\rule{3.132pt}{0.800pt}}
\multiput(1026.00,361.34)(6.500,-3.000){2}{\rule{1.566pt}{0.800pt}}
\put(1039,356.84){\rule{3.132pt}{0.800pt}}
\multiput(1039.00,358.34)(6.500,-3.000){2}{\rule{1.566pt}{0.800pt}}
\put(1052,353.84){\rule{3.132pt}{0.800pt}}
\multiput(1052.00,355.34)(6.500,-3.000){2}{\rule{1.566pt}{0.800pt}}
\put(1065,351.34){\rule{3.132pt}{0.800pt}}
\multiput(1065.00,352.34)(6.500,-2.000){2}{\rule{1.566pt}{0.800pt}}
\put(1078,348.84){\rule{2.891pt}{0.800pt}}
\multiput(1078.00,350.34)(6.000,-3.000){2}{\rule{1.445pt}{0.800pt}}
\put(1090,345.84){\rule{3.132pt}{0.800pt}}
\multiput(1090.00,347.34)(6.500,-3.000){2}{\rule{1.566pt}{0.800pt}}
\put(1103,343.34){\rule{3.132pt}{0.800pt}}
\multiput(1103.00,344.34)(6.500,-2.000){2}{\rule{1.566pt}{0.800pt}}
\put(1116,340.84){\rule{3.132pt}{0.800pt}}
\multiput(1116.00,342.34)(6.500,-3.000){2}{\rule{1.566pt}{0.800pt}}
\put(1129,338.34){\rule{3.132pt}{0.800pt}}
\multiput(1129.00,339.34)(6.500,-2.000){2}{\rule{1.566pt}{0.800pt}}
\put(1142,335.84){\rule{3.132pt}{0.800pt}}
\multiput(1142.00,337.34)(6.500,-3.000){2}{\rule{1.566pt}{0.800pt}}
\put(1155,333.34){\rule{3.132pt}{0.800pt}}
\multiput(1155.00,334.34)(6.500,-2.000){2}{\rule{1.566pt}{0.800pt}}
\put(1168,331.34){\rule{3.132pt}{0.800pt}}
\multiput(1168.00,332.34)(6.500,-2.000){2}{\rule{1.566pt}{0.800pt}}
\put(1181,328.84){\rule{3.132pt}{0.800pt}}
\multiput(1181.00,330.34)(6.500,-3.000){2}{\rule{1.566pt}{0.800pt}}
\put(1194,326.34){\rule{3.132pt}{0.800pt}}
\multiput(1194.00,327.34)(6.500,-2.000){2}{\rule{1.566pt}{0.800pt}}
\put(1207,324.34){\rule{3.132pt}{0.800pt}}
\multiput(1207.00,325.34)(6.500,-2.000){2}{\rule{1.566pt}{0.800pt}}
\put(1220,322.34){\rule{2.891pt}{0.800pt}}
\multiput(1220.00,323.34)(6.000,-2.000){2}{\rule{1.445pt}{0.800pt}}
\put(1232,320.34){\rule{3.132pt}{0.800pt}}
\multiput(1232.00,321.34)(6.500,-2.000){2}{\rule{1.566pt}{0.800pt}}
\put(1245,318.34){\rule{3.132pt}{0.800pt}}
\multiput(1245.00,319.34)(6.500,-2.000){2}{\rule{1.566pt}{0.800pt}}
\put(1258,316.34){\rule{3.132pt}{0.800pt}}
\multiput(1258.00,317.34)(6.500,-2.000){2}{\rule{1.566pt}{0.800pt}}
\put(1271,314.34){\rule{3.132pt}{0.800pt}}
\multiput(1271.00,315.34)(6.500,-2.000){2}{\rule{1.566pt}{0.800pt}}
\put(1284,312.34){\rule{3.132pt}{0.800pt}}
\multiput(1284.00,313.34)(6.500,-2.000){2}{\rule{1.566pt}{0.800pt}}
\put(1297,310.34){\rule{3.132pt}{0.800pt}}
\multiput(1297.00,311.34)(6.500,-2.000){2}{\rule{1.566pt}{0.800pt}}
\put(1310,308.34){\rule{3.132pt}{0.800pt}}
\multiput(1310.00,309.34)(6.500,-2.000){2}{\rule{1.566pt}{0.800pt}}
\put(1323,306.34){\rule{3.132pt}{0.800pt}}
\multiput(1323.00,307.34)(6.500,-2.000){2}{\rule{1.566pt}{0.800pt}}
\put(1336,304.84){\rule{3.132pt}{0.800pt}}
\multiput(1336.00,305.34)(6.500,-1.000){2}{\rule{1.566pt}{0.800pt}}
\put(1349,303.34){\rule{3.132pt}{0.800pt}}
\multiput(1349.00,304.34)(6.500,-2.000){2}{\rule{1.566pt}{0.800pt}}
\put(1362,301.34){\rule{2.891pt}{0.800pt}}
\multiput(1362.00,302.34)(6.000,-2.000){2}{\rule{1.445pt}{0.800pt}}
\put(1374,299.84){\rule{3.132pt}{0.800pt}}
\multiput(1374.00,300.34)(6.500,-1.000){2}{\rule{1.566pt}{0.800pt}}
\put(1387,298.34){\rule{3.132pt}{0.800pt}}
\multiput(1387.00,299.34)(6.500,-2.000){2}{\rule{1.566pt}{0.800pt}}
\put(1400,296.34){\rule{3.132pt}{0.800pt}}
\multiput(1400.00,297.34)(6.500,-2.000){2}{\rule{1.566pt}{0.800pt}}
\put(1413,294.84){\rule{3.132pt}{0.800pt}}
\multiput(1413.00,295.34)(6.500,-1.000){2}{\rule{1.566pt}{0.800pt}}
\put(1426,293.34){\rule{3.132pt}{0.800pt}}
\multiput(1426.00,294.34)(6.500,-2.000){2}{\rule{1.566pt}{0.800pt}}
\end{picture}

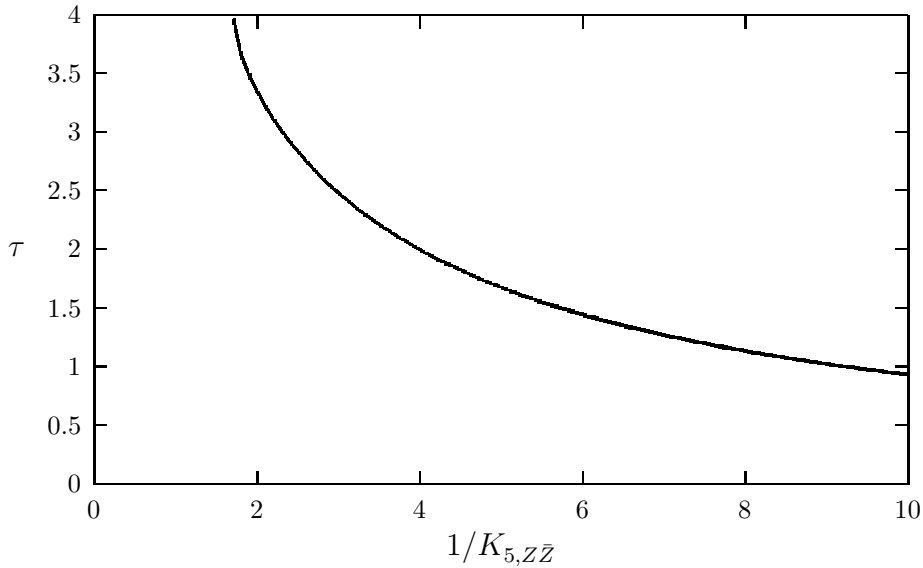
Fig.2: The solution $m_\Phi^2=0$ for $\tan \theta =1$.
\end{center}

\section{Conclusion}
We have studied on the flatness condition within the framwework 
of M-theory with and without five-brane.
In the case without five-brane, we can not obtain  a realistic solution, 
because the condition (\ref{const1}) constrains severely.
In the case with five-brane, the $F$-terms of five-brane moduli fields 
can also contribute the supersymmetry breaking and the vacuum energy.
We have shown such effects are important to obtain the realistic 
solutions for $m_\Phi^2=0$.
In particular, a large value of $1/K_{5,Z \bar Z}$ is favorable.
In addition, the condition (\ref{const1}) is relaxed because of 
five-brane effects.

\section*{Acknowledgement} 
The authors would like to thank J.~Kubo and M.~Roos
for their encouragements and useful discussions.
The research of T.K. was partially supported 
by the Academy of Finland (no. 44129).
S.M.M. is indebted to the Magnus Ehrnrooths Foundation for its support.

\end{document}